\documentclass[showpacs,twocolumn,aps]{revtex4}
\usepackage{graphicx}
\usepackage{dcolumn}
\usepackage{bm}
\usepackage{color}
\usepackage[squaren]{SIunits}
\usepackage{amssymb}
\usepackage{natbib}

\begin{document}

\newcommand{\ie}{{\it i.e.}}
\newcommand{\eg}{{\it e.g.}}
\newcommand{\etal}{{\it et al.}}


\title{Crystal structure, physical properties and superconductivity in $A_{x}$Fe$_2$Se$_2$ single crystals}

\author{X. G. Luo, X. F. Wang, J. J. Ying, Y. J. Yan, Z. Y. Li, M. Zhang, A. F. Wang, P. Cheng, Z. J. Xiang, G. J. Ye, R. H. Liu and X. H. Chen}
\altaffiliation{E-mail: chenxh@ustc.edu.cn\\ } \affiliation{Hefei
National Laboratory for Physical Science at Microscale and
Department of Physics, University of Science and Technology of
China, Hefei, Anhui 230026, People's Republic of China}

\date{\today}


\begin{abstract}

We studied the correlation among structure and transport properties
and superconductivity in the different $A_x$Fe$_2$Se$_2$ single crystals
($A$ = K, Rb, and Cs). Two sets of (00$l$) reflections are observed in the
X-ray single crystal diffraction patterns, and arise from the
intrinsic inhomogeneous distribution of the intercalated alkali
atoms. The occurrence of superconductivity is closely related to the
{\sl c}-axis lattice constant, and the $A$ content is crucial to
superconductivity. The hump observed in resistivity seems to be
irrelevant to superconductivity. There exist many deficiencies within the FeSe layers in $A_x$Fe$_2$Se$_2$, while their $T_{\rm c}$ does not change so much. In this sense, superconductivity is robust to the Fe and Se
vacancies. Very high resistivity in the normal state should arise from
such defects in the conducting FeSe layers. $A_x$Fe$_2$Se$_2$ ($A$ = K, Rb, and Cs) single crystals show the same susceptibility behavior in the normal state, and no anomaly is observed in susceptibility at the hump
temperature in resistivity. The clear jump in specific heat for
Rb$_x$Fe$_2$Se$_2$ and K$_x$Fe$_2$Se$_2$ single crystals shows the good
bulk superconductivity in these crystals.

\end{abstract}


\maketitle

The newly discovered iron-based superconductors have attracted
worldwide attention in the past three
years \cite{Kamihara,chenxh,ZARen,rotter1,Liu} because of their high
superconducting transition temperature ($T_{\rm c}$ as high as 55 K)
and the fact that superconductivity emerges proximity to
the magnetically ordered state. \cite{PCDai,hchen}, which were considered
to be taken as a comparison with the superconducting cuprates for
finding out the mechanism of the high-$T_{\rm c}$ superconductivity. The
highest $T_{\rm c}$ of the FeAs-based pnictides reaches 55 K at ambient
pressure, while the anti-PbO type FeSe$_x$, owning the extremely simple
structure with the edge-sharing FeSe$_4$ tetrahedra formed FeSe
layers stacking along the {\sl c}-axis, displays a lower $T_{\rm c}$ of
8 K at ambient pressure \cite{MKWu}. After $T_{\rm c}$ was enhanced
to as high as 37 K by applying high pressure \cite{cava}, the efforts of
introducing structures between the FeSe layers, just like the FeAs-analogues
being, successfully induced ~30 K superconductivity by intercalating
the alkali (K, Rb, Cs) and Tl atoms into between the FeSe
layers \cite{xlchen,Mizuguchi,Wang,Ying,Krzton,Fang}. The
intercalated FeSe superconductors show some distinct physical
properties from the FeAs-based superconductors, such as,
superconductivity with very high normal-state resistivity, a broad
hump in resistivity. The Fe content is important in controlling the
magnetic and superconducting properties in the iron chalcogenides,
additional iron would greatly affect its structural and magnetic
properties \cite{Bao}, and superconductivity can be enhanced by
the de-intercalation of the interstitial iron \cite{Bendele, Rodriguez}.
However, vacancies have been found to be able to exist at either $A$
site, or within the conducting FeSe layers in chemical formula
$A_x$Fe$_2$Se$_2$, and it is much more complicated than that in
the Fe-Se systems. The intercalated alkali atoms could be crucial to the
superconductivity. The normal-state resistivity should be influenced
by these vacancies seriously. But how these vacancies affect the
physical properties still remains unresolved.

In this article, we systematically studied the effect of the starting
materials and the heating process on the single crystal growth for
$A_x$Fe$_2$Se$_2$ ($A$ = K, Rb, and Cs), and measured the physical
properties of these single crystals and determined their crystal
structures. It is found that two sets of (00$l$) reflections exist
in all the crystals, and superconductivity is closely related to the
{\sl c}-axis lattice constant, indicating that the $A$ content is
crucial to the superconductivity. The hump in resistivity arises
from the defects within the conducting FeSe layers and is irrelevant to superconductivity.
No anomaly is observed in magnetic susceptibility at the temperature
of hump in resistivity. The clear jump in specific heat for
superconducting K$_x$Fe$_2$Se$_2$ and Rb$_x$Fe$_2$Se$_2$ single
crystals indicates the good bulk superconductivity in these crystals.

Single crystals $A_x$Fe$_2$Se$_2$ ($A$ = K, Rb, and Cs) were grown by Bridgeman
method as described elsewhere\cite{Ying,Wang}. The starting
materials and the heating process are very important to get
superconducting single crystal, and even is slightly changed to
dramatically affect its physical properties. The three different
batches of Rb$_x$Fe$_2$Se$_2$ and three batches of K$_x$Fe$_2$Se$_2$
single crystals were gotten by the slightly change of the heating
temperatures and starting materials. The single crystals were
characterized by X-ray single crystal diffraction, magnetic
susceptibility, and electrical transport measurements. X-ray single
crystal diffraction was performed on a TTRAX3 theta/theta rotating
anode X-ray Diffractometer (Japan) with Cu K$\alpha$ radiation and a
fixed graphite monochromator. Magnetic susceptibility measurements
were carried out using the {\sl Quantum Design} MPMS-SQUID. The
measurement of resistivity and magnetoresistance were done on the
{\sl Quantum Design} PPMS-9.

\begin{figure}[t]
\includegraphics[width = 0.48\textwidth]{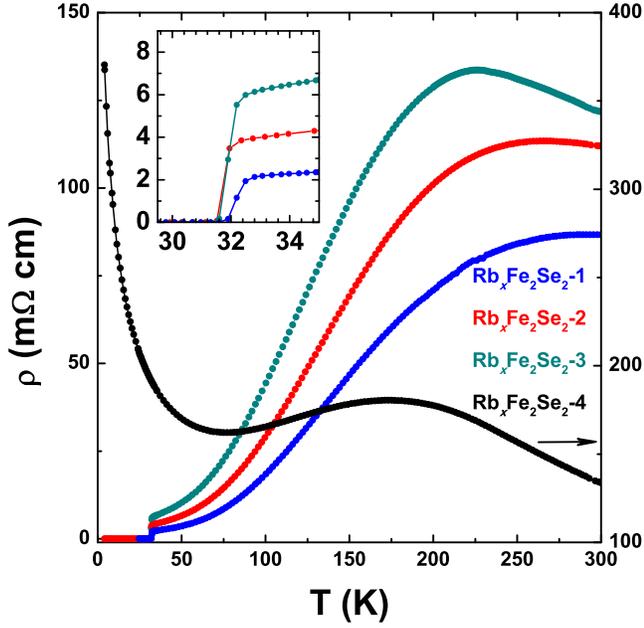}
\caption{(Color online) Temperature dependence of resistivity for
the three batches of Rb$_x$Fe$_2$Se$_2$ single crystals:
Rb$_x$Fe$_2$Se$_2$-1 (blue line), Rb$_x$Fe$_2$Se$_2$-2 (red line),
Rb$_x$Fe$_2$Se$_2$-3 (dark cyan line) and Rb$_x$Fe$_2$Se$_2$-4
(black line). The inset is zoomed plot of (b) around $T_{\rm c}$.
}
\end{figure}

The typical temperature dependence of resistivity is observed for
three batches of Rb$_x$Fe$_2$Se$_2$ single crystals as shown in
Fig.1. Among these crystals, Rb$_x$Fe$_2$Se$_2$-1 was obtained with
nominal composition as Rb$_{0.8}$Fe$_2$Se$_{1.96}$ by being melt at
1080$\celsius$ and turning off furnace at 950 $\celsius$;
Rb$_x$Fe$_2$Se$_2$-2 and Rb$_x$Fe$_2$Se$_2$-3 came from the same
batch with the nominal composition as Rb$_{0.8}$Fe$_2$Se$_{2}$ by
being melt at 1030$\celsius$ and turning off furnace at 700
$\celsius$; Rb$_x$Fe$_2$Se$_2$-4 was grown with nominal composition
as Rb$_{0.8}$Fe$_2$Se$_{1.96}$ by being melt at 1030$\celsius$ and
switching off furnace at 700 $\celsius$. The resistivity of
Rb$_x$Fe$_2$Se$_2$-1 shows a very small hump around 290 K and then
becomes metallic below this temperature with the residual
resistivity ratio RRR = $R$(300 K)/$R$(35 K) $\approx$ 37.2.
Superconductivity appears below 32.4 K and zero resistance is
reached at 31.9 K. The superconducting transition width of
Rb$_x$Fe$_2$Se$_2$-1 is as narrow as 0.5 K although the resistivity
at room temperature is as large as 70 m$\Omega$ cm at 300 K. The
hump temperature of resistivity ($T_{\rm hump}$) shifts to 265 K and
225 K for Rb$_x$Fe$_2$Se$_2$-2 and Rb$_x$Fe$_2$Se$_2$-3,
respectively. The RRR decreases to 26.1 and 17.3 for the two
crystals, respectively. These results indicate that the metallicity
of Rb$_x$Fe$_2$Se$_2$-2 and Rb$_x$Fe$_2$Se$_2$-3 is weaker compared
to Rb$_x$Fe$_2$Se$_2$-1. However, as we can see from the inset of Fig.1
and Table I, the superconducting transition temperature seems not to vary with
the change in the resistivity behavior and their values. The onset
and zero resistance temperature for Rb$_x$Fe$_2$Se$_2$-2 and
Rb$_x$Fe$_2$Se$_2$-3 are 32.0, 32.4 and 31.5, 31.6 K, respectively.
Thus one can see that the humps in resistivity seems to be
irrelevant to superconductivity. The large magnitude of the
normal-state resistivity compared to FeAs-base pnictides \cite{XFWang,HaihWen,HaihWen1} and FeSe \cite{Braithwaite} (usually with $\rho$(300 K) much less than 1 m$\Omega$ cm)
reflects the existence of many deficiencies within the
conducting FeSe layers in these Rb$_x$Fe$_2$Se$_2$ single crystals.
The hump in resistivity should arise from such large amount of
defects within the conducting FeSe layers. In this
sense, superconductivity is quite robust to the vacancies within the FeSe layers.
For Rb$_x$Fe$_2$Se$_2$-4, although the resistivity still shows a
hump at around 170 K, no superconductivity can be observed, and a
strong semiconducting/insulator-like behavior is observed below 70
K.

\begin{figure}[t]
\includegraphics[width = 0.48\textwidth]{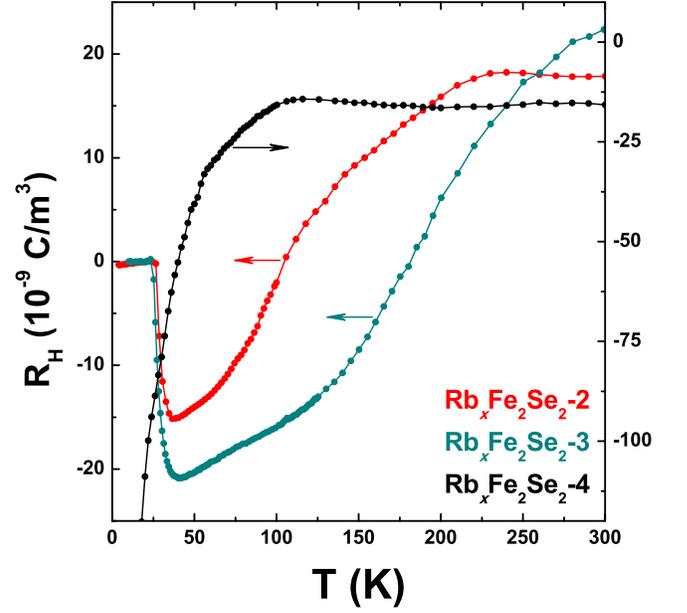}
\caption{(Color online) Hall coefficient as a function of
temperature for single crystals: Rb$_x$Fe$_2$Se$_2$-2,
Rb$_x$Fe$_2$Se$_2$-3 and Rb$_x$Fe$_2$Se$_2$-4.}
\end{figure}

For comparison, we measured the temperature dependence of the Hall
coefficient on the exactly same pieces of Rb$_x$Fe$_2$Se$_2$-2,
Rb$_x$Fe$_2$Se$_2$-3 and Rb$_x$Fe$_2$Se$_2$-4 as shown in Fig.2. The
Hall coefficient of the superconducting crystals Rb$_x$Fe$_2$Se$_2$-2
and Rb$_x$Fe$_2$Se$_2$-3 is positive at high temperature, and
gradually decreases with decreasing temperature and then becomes negative
at low temperature. The sign change of the Hall coefficient is also
observed in superconducting Tl$_{0.58}$Rb$_{0.42}$Fe$_{1.72}$Se$_2$
samples previously \cite{Wang1}. Actually from ARPES results in $A_x$Fe$_2$Se$_2$ ($A$ =
K and Cs) \cite{Zhang,DingH}, the $A_x$Fe$_2$Se$_2$ are electron
over-doped and only electron pockets can be observed. However, hole
pockets could exist in the superconducting samples due to sign
change of Hall coefficient based on the Hall effect measurements,
suggesting a possibly multi-band nature of the superconductivity. For the
sample without superconductivity, the Hall coefficient is negative
in the whole temperature range. It indicates that dominant carrier
is electron for non-superconducting crystal. It seems that the hole
pocket might be quite important for the superconductivity.
Therefore, it needs to be further investigated. What's more, no
anomaly in the Hall coefficient is observed at $T_{\rm hump}$. It
suggests that the humps in resistivity for $A_x$Fe$_2$Se$_2$ crystals are not
related to a structural or magnetic transition, being contrasting to
the facts that the anomaly of resistivity in underdoped FeAs-based
superconductor is always relevant to a structural/magnetic
transition.

\begin{figure}[t]
\includegraphics[width = 0.48\textwidth]{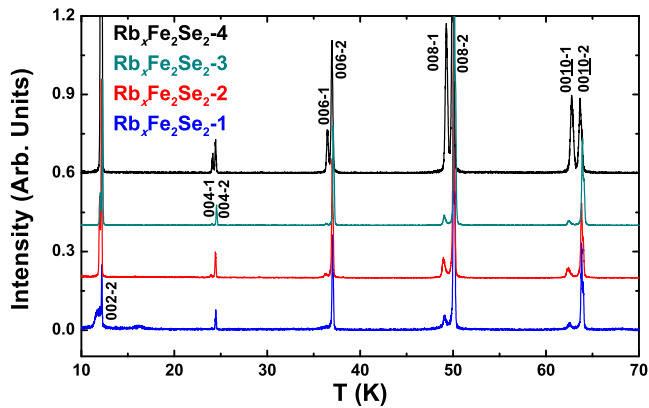}
\caption{(Color online) The X-ray single crystal diffraction
patterns for the different batches Rb$_x$Fe$_2$Se$_2$ single
crystals.}
\end{figure}

As shown in Fig. 1, one can observe almost the same $T_{\rm c}$ for
the single crystals with the different $T_{\rm hump}$. The X-ray single
crystal diffraction was carried out for the same four pieces
Rb$_x$Fe$_2$Se$_2$ single crystals shown in Fig. 1 to find out the
relationship of superconductivity and structure. The X-ray
diffraction (XRD) patterns are shown in Fig. 3. Surprisingly, two
sets of (00$l$) reflections are observed in all the four samples.
The two {\sl c}-axis lattice parameters $c$1 and $c$2 are obtained
(listed in Table I). The {\sl c}-axis lattice parameters $c$1 and $c$2
correspond to the two sets of reflections with weak and strong
intensities, respectively. These two distinct sets of reflections
could arise from the inhomogeneous distribution of the intercalated
Rb atoms. Considering the fact that the superconducting crystals
show nearly fully shielding fraction, the reflections with $c$2
should be responsible for the superconductivity. From the
superconducting to non-superconducting crystal, $c$1 is reduced by
0.55$\%$ while $c$2 is enhanced by more than 0.14$\%$. It is found
that the insulator-like behavior is enhanced with losing
superconductivity. It indicates that superconductivity may exist
within a limited range of the {\sl c}-axis lattice parameter. In other word,
the Rb content is crucial to the occurrence of superconductivity because
the {\sl c}-axis lattice parameter strongly depends on the Rb content.

\begin{table*}[t]
\tabcolsep 0pt \caption{The {\sl c}-axis lattice parameters $c$1 and
$c$2 corresponding to the two sets of reflections with weak and
strong intensities, respectively. $T_{\rm c}^{\rm zero}$, $T_{\rm
c}^{\rm onset}$, the hump temperature in resistivity ($T_{\rm
hump}$) and residual resistivity ratio (RRR = $R$(300 K) / $R$(35
K)) for all the crystals $A_{x}$Fe$_2$Se$_2$ ($A$=K, Rb, Cs).}
\vspace*{-12pt}
\begin{center}
\def\temptablewidth{1\textwidth}
{\rule{\temptablewidth}{1pt}}
\begin{tabular*}{\temptablewidth}{@{\extracolsep{\fill}}ccccccc}
 sample name & $c$1 (${\rm \AA}$) & $c$2 (${\rm \AA}$)&$T_{\rm c}^{\rm zero}$ (K) &$T_{\rm c}^{\rm onset}$ (K) & $T_{\rm hump}$ (K) & RRR\\   \hline
      Rb$_x$Fe$_2$Se$_2$-1 & 14.873 & 14.569 & 31.9 & 32.4 & 290 &37.2\\
      Rb$_x$Fe$_2$Se$_2$-2 & 14.873 & 14.582 & 31.5 & 32.0 & 265 &26.1\\
      Rb$_x$Fe$_2$Se$_2$-3 & 14.874 & 14.574 & 31.6 & 32.4 & 225 &17.3\\
      Rb$_x$Fe$_2$Se$_2$-4 & 14.792 & 14.604 & & & 170  \\ \hline
      K$_x$Fe$_2$Se$_2$-1 & 14.292 & 14.086 & 31.2 & 31.7 & 220 &21.2 \\
      K$_x$Fe$_2$Se$_2$-2 & 14.282 & 14.062 & 29.2 & 30.8 & 120 &0.65 \\
      K$_x$Fe$_2$Se$_2$-3 & 14.201 & 14.107 & & & 160  \\    \hline
      Cs$_x$Fe$_2$Se$_2$ & 15.556 & 15.285 & 28.3 & 30.3 & no hump $<$300 K &16.3  \\
       \end{tabular*}
       {\rule{\temptablewidth}{1pt}}
\end{center}
\end{table*}

\begin{figure}[th]
\includegraphics[width = 0.45\textwidth]{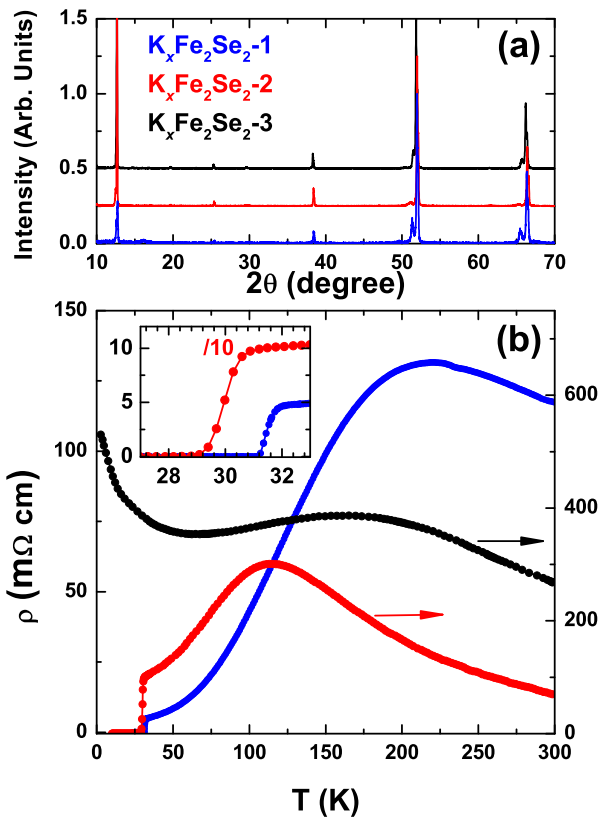}
\caption{(Color online) (a): The X-ray single crystal diffraction
patterns for the different batches K$_x$Fe$_2$Se$_2$ single
crystals: K$_x$Fe$_2$Se$_2$-1 (blue line), K$_x$Fe$_2$Se$_2$-2 (red
line) and K$_x$Fe$_2$Se$_2$-3 (black line); (b): The temperature
dependence of the resistivity of these three different
K$_x$Fe$_2$Se$_2$ single crystals. The inset is zoomed plot of (b)
around $T_{\rm c}$. }
\end{figure}

\begin{figure}[th]
\includegraphics[width = 0.45\textwidth]{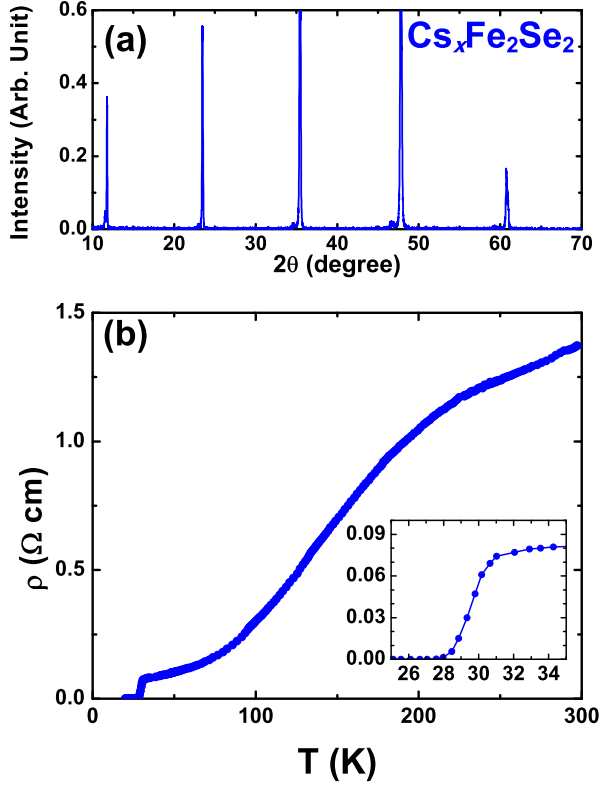}
\caption{(Color online) (a): The X-ray single crystal diffraction
pattern for the Cs$_x$Fe$_2$Se$_2$ single crystal; (b): The
temperature dependence of the resistivity of the Cs$_x$Fe$_2$Se$_2$
single crystal. The inset is zoomed plot of (b) around $T_{\rm c}$.
}
\end{figure}

We then carefully measured resistivity and the XRD patterns for three pieces of K$_x$Fe$_2$Se$_2$ crystals, as shown in Fig. 4 . It shows that the
K$_x$Fe$_2$Se$_2$ crystals exhibit the obviously different resistivity
behavior. K$_x$Fe$_2$Se$_2$-1 was grown using K$_{0.8}$(FeSe)$_2$ as
starting materials and being melt at 1030 $\celsius$ for 3 hours.
K$_x$Fe$_2$Se$_2$-2 and K$_x$Fe$_2$Se$_2$-3 were grown by using
K$_{0.8}$(FeSe)$_2$ as starting materials and being melt at 1030
$\celsius$ for 2 hours and 950 $\celsius$ for 20 hours,
respectively. In Fig.4a, two sets of (00$l$) reflections can be
observed, and this behavior is the same with those in the
Rb$_x$Fe$_2$Se$_2$ samples, suggesting that the inhomogeneous
distribution of the intercalated alkali atoms in the crystals is
common feature. For the non-superconducting sample, the {\sl c}-axis
lattice constant $c$1 is smaller by more than 0.64$\%$, while $c$2
is larger by 0.15$\%$ than those in the superconducting samples.
These results are consistent with the results observed in the
Rb$_x$Fe$_2$Se$_2$ samples, indicating that the content of alkali
atom plays a crucial role for the occurrence of superconductivity.
K$_x$Fe$_2$Se$_2$-1 shows a broad resistivity hump at about 220 K
and superconductivity at 31.7 K. For the K$_x$Fe$_2$Se$_2$-2,
$T_{\rm hump}$ shifts to 120 K and superconductivity shows up at
30.3 K. Although the superconductivity disappears in
K$_x$Fe$_2$Se$_2$-3, the $T_{\rm hump}$ for K$_x$Fe$_2$Se$_2$-3 is
higher than that for the superconducting K$_x$Fe$_2$Se$_2$-2,
strongly demonstrating that the superconductivity is not correlated
to the hump in resistivity. The position of the hump in resistivity
reflects the vacancy level within the conducting FeSe
layers. It suggests that the vacancies within the FeSe layers have much weaker
correlation to superconductivity than the content of intercalated
alkali atoms does.

\begin{figure}[th]
\includegraphics[width = 0.45\textwidth]{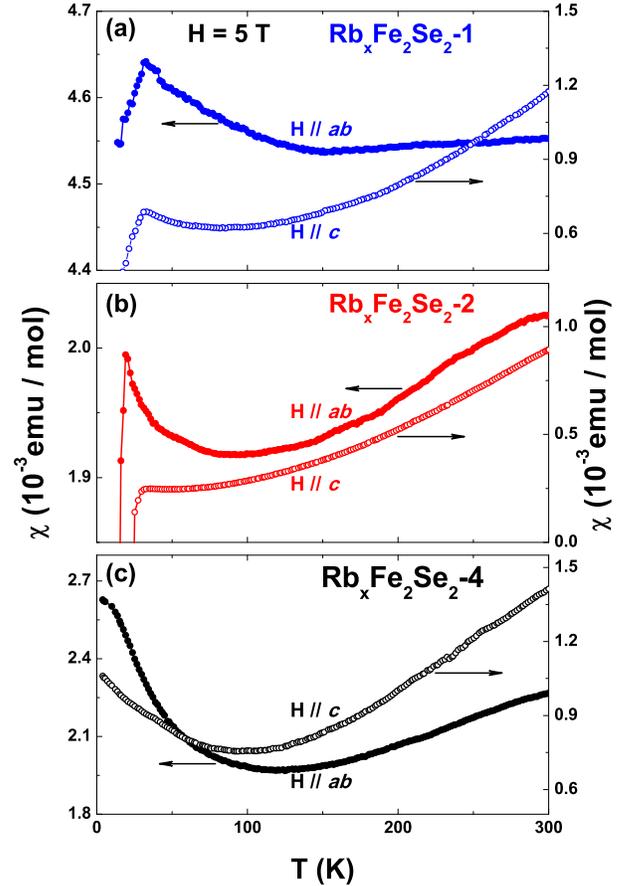}
\caption{(Color online) The temperature dependence of the
susceptibility with the magnetic field of 5T applied parallel and
perpendicular to the {\sl c}-axis for (a): Rb$_x$Fe$_2$Se$_2$-1;
(b): Rb$_x$Fe$_2$Se$_2$-2; (c): Rb$_x$Fe$_2$Se$_2$-4. }
\end{figure}

Figure 5 shows the X-ray single crystal diffraction pattern and the
temperature dependence of resistivity for Cs$_x$Fe$_2$Se$_2$ single
crystal. Totally metallic resistivity can be observed below 300 K.
Superconductivity was observed with $T_{\rm c}^{\rm onset}$ = 30.3 K
and $T_{\rm c}^{\rm zero}$ = 28.3 K. In the XRD patterns of Fig.5a,
very small reflections corresponding to $c$1 can still be found
except for the main reflections with $c$2. It suggests the
inhomogeneous distribution of alkali atoms is common for all the
$A_x$Fe$_2$Se$_2$ single crystals. It is worthy to note that the
superconductivity always shows up around 30 K for the crystals
A$_x$Fe$_2$Se$_2$ with changing the intercalated alkali atom $A$ from
K, Rb to Cs. The $T_{\rm c}$ seems not to depend on the ionic radii
of the intercalated alkali atoms although the superconductivity
strongly depends on the $A$ content. As shown in Fig.1 and Fig.4b,
the hump in resistivity changes pronouncedly for the same alkali
atom case, while the $T_c$ is nearly the same (about 30 K). Very
large normal-state resistivity is observed in all the above
$A_x$Fe$_2$Se$_2$ single crystals, suggesting that large amount of
deficiencies within the conducting FeSe layers for all of
these crystals. Based on these observations, $T_{\rm c}$ seems to be robust to the vacancies within the FeSe layers.

\begin{figure}[th]
\includegraphics[width = 0.45\textwidth]{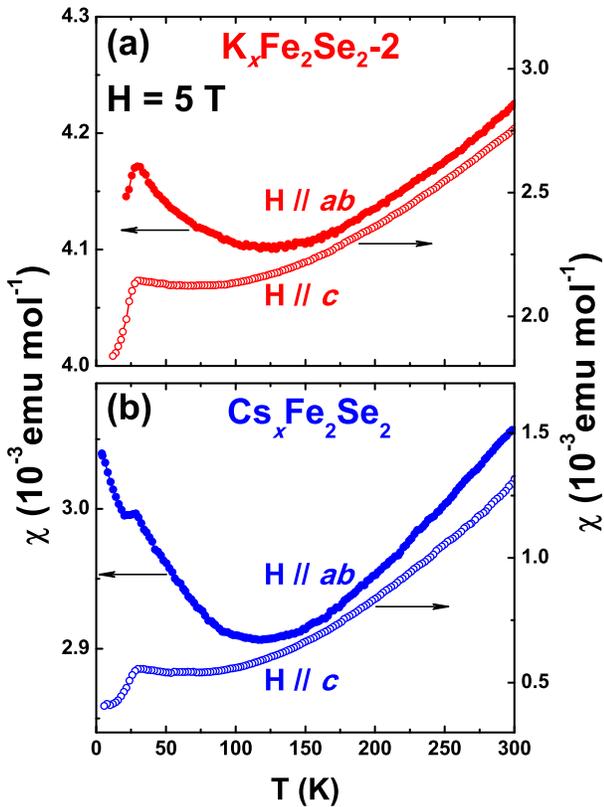}
\caption{(Color online) The temperature dependence of the
susceptibility with the magnetic field of 5 T applied parallel and
perpendicular to the {\sl c}-axis for (a): K$_x$Fe$_2$Se$_2$-2; and
(b): Cs$_x$Fe$_2$Se$_2$.}
\end{figure}

Magnetic susceptibility was measured on the $A_x$Fe$_2$Se$_2$ single
crystals to investigate the correlation between the normal-state
resistivity and magnetism. Figure 6 shows the anisotropic magnetic
susceptibility with the magnetic field of 5 T applied within the {\sl
ab}-plane and along the {\sl c}-axis for Rb$_x$Fe$_2$Se$_2$-1,
Rb$_x$Fe$_2$Se$_2$-2, Rb$_x$Fe$_2$Se$_2$-4, respectively. Although
the samples show very different resistivity behavior, such as
the different $T_{\rm hump}$s, the magnitude of resistivity and $T_c$, the
normal-state susceptibility shows the quite similar behavior to each
other. As the field is applied within the {\sl ab}-plane, the magnitude of
susceptibility varies within 20$\%$ in the normal state and
the susceptibility itself shows a broad minimum. No anomaly can be found
at $T_{\rm hump}$, suggesting that hump in resistivity cannot
be ascribed to a magnetic transition. Fig.7 shows the susceptibility
with the magnetic field of 5 T applied parallel and perpendicular to
the {\sl c}-axis for K$_x$Fe$_2$Se$_2$-2 and Cs$_x$Fe$_2$Se$_2$,
respectively. The similar behavior with those of Rb$_x$Fe$_2$Se$_2$
crystals shown in Fig.6 is observed. These results indicate that
although the electronic properties change dramatically from system
to system and from crystal to crystal, the magnetic property does
not change a lot.

\begin{figure}[th]
\includegraphics[width = 0.45\textwidth]{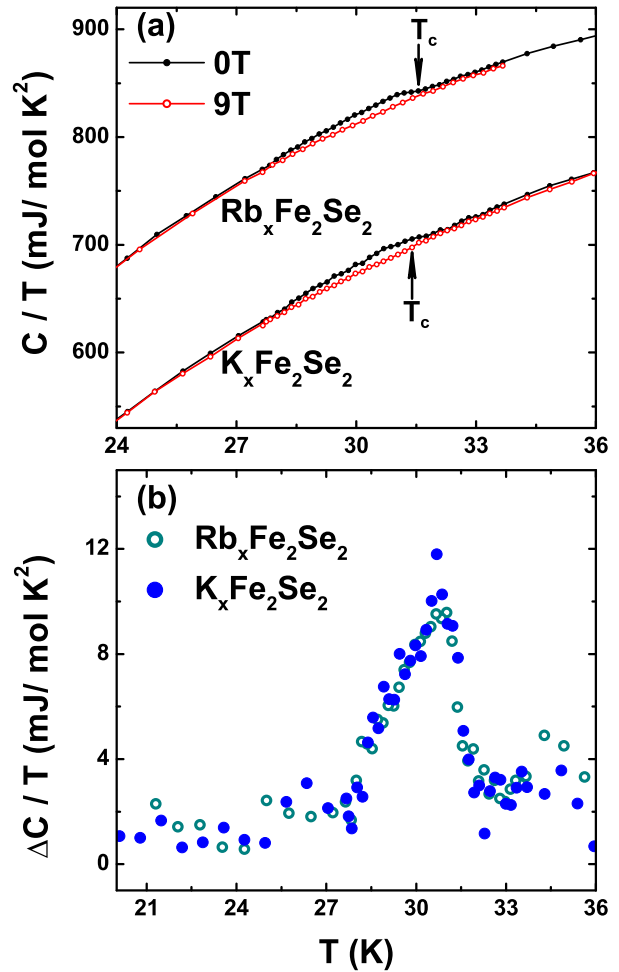}
\caption{(Color online) (a): The heat capacity as a function of
temperature for Rb$_x$Fe$_2$Se$_2$-2 and K$_x$Fe$_2$Se$_2$-1 at the
magnetic field of 0T and 9T applied along {\sl c}-axis. (b): The
heat capacity difference between 0T and 9T for Rb$_x$Fe$_2$Se$_2$-2
and K$_x$Fe$_2$Se$_2$-1. A clear heat peak was observed, indicating
the good bulk superconductivity. }
\end{figure}

Fig.8a shows the temperature dependence of the specific heat ($C/T$)
around $T_{\rm c}$ at the magnetic field of 0 T and 9 T for
Rb$_x$Fe$_2$Se$_2$-2 and K$_x$Fe$_2$Se$_2$-1, respectively. At 0 T,
one can see the clear specific heat anomaly at $T_{\rm c}$. Although 9 T
is far lower than the upper critical field, which is estimated
higher than 100 T,\cite{Ying,Wang} the anomaly in specific heat is
completely suppressed. The specific heat jump ($C$(0T)-$C$(9T))/$T$
against $T$ is plotted in Fig.8b for the two crystals. The heat
capacity jumps for Rb$_x$Fe$_2$Se$_2$-2 and K$_x$Fe$_2$Se$_2$-1
crystals shows almost the same behavior. The clear heat capacity
jump in the superconducting samples definitely indicates the good bulk
superconductivity in these crystals.

The X-ray single crystal diffraction patterns reveals two sets of
(00$l$) reflections existing in all the crystals of A$_x$Fe$_2$Se$_2$.
Such two sets of reflections strongly depend on the starting
composition and heating process. Although the superconducting phase
is dominant, the trace of second phase is still observed as shown in
Fig.5a. These results indicate the existence of inhomogeneous
distribution of the $A$ atoms in all the crystals. It is found that
superconductivity is closely related to the {\sl c}-axis lattice
constant, indicating that the $A$ content is crucial to the
superconductivity because the {\sl c}-axis lattice parameter strongly
depends on the $A$ content. The $A$ content in single crystals is quite
sensitive to the nominal composition and condition of crystal
growth. Therefore, it is not easy to grow the single crystal with
superconductivity. The very large normal-state resistivity relative
to other iron pnictide superconductors suggests the large amount of deficiencies within the conducting FeSe layers for all the
$A_x$Fe$_2$Se$_2$ single crystals. The hump in resistivity should
arise from such defects, and seems to be irrelevant to
superconductivity. Despite of the existence of many deficiencies within the conducting FeSe layers in $A_x$Fe$_2$Se$_2$, $T_c$ does not change a lot with varying $A$ from K to Rb and Cs. Therefore, superconductivity seems
robust to the such vacancies. No anomaly is observed in
magnetic susceptibility at the hump temperature in resistivity.
It suggests that the humps in resistivity in $A_x$Fe$_2$Se$_2$ are
not related to a structural or magnetic transition, being
contrasting to the facts that the anomaly of resistivity in the
underdoped FeAs-based superconductor is always relevant to a
structural/magnetic transition.

In conclusion, we systematically studied the structure by the X-ray
single crystal diffraction and measured the transport properties in
the $A_x$Fe$_2$Se$_2$ single crystals. All the samples show two sets
of (00$l$) reflections in X-ray single crystal diffraction patterns,
indicating the intrinsically inhomogeneous distribution of the
intercalated $A$ atoms. The occurrence of Superconductivity is closely
related to the {\sl c}-axis lattice parameter, indicating that the $A$
content is crucial to the superconductivity in $A_{x}$Fe$_2$Se$_2$.
The very large magnitude of the normal-state resistivity reflects
the large amount of deficiencies within the conducting FeSe layers.
The hump in resistivity should originate from these defects and is found to be irrelevant to the superconductivity. In
this sense, superconductivity is robust to the vacancies within the FeSe layers.
No anomaly in susceptibility is observed to be associated with the
hump in resistivity. The clear jump in specific heat for
Rb$_x$Fe$_2$Se$_2$ and K$_x$Fe$_2$Se$_2$ superconducting single
crystals indicates the good bulk superconductivity in these crystals.

{\bf ACKNOWLEDGEMENT} This work is supported by the Natural Science
Foundation of China and by the Ministry of Science and Technology of China, and by Chinese Academy of Sciences.\\


\begin{thebibliography}{}


\bibitem{Kamihara}
Yoichi Kamihara, Takumi Watanabe, Masahiro Hirano and Hideo Hosono,
J. Am. Chem. Soc. {\bf 130}, 3296 (2008).

\bibitem{chenxh}
X. H. Chen, T. Wu, G. Wu, R. H. Liu, H. Chen and D. F. Fang, Nature
{\bf 453}, 761(2008).

\bibitem{ZARen}
Z. A. Ren, W. Lu, J. Yang, W. Yi, X. L. Shen, Z. C. Li, G. C. Che,
X. L. Dong, L. L. Sun, F. Zhou and Z. X. Zhao, Chin. Phys. Lett. {\bf 25} , 2215(2008).

\bibitem{rotter1}
M. Rotter, M. Tegel, D. Johrendt, Phys. Rev. Lett. {\bf 101},
107006(2008).

\bibitem{Liu}
R. H. Liu, G.Wu, T. Wu, D. F. Fang, H. Chen, S. Y. Li, K. Liu, Y. L.
Xie, X. F.Wang, R. L. Yang, L. Ding, C. He, D. L. Feng and X. H.
Chen, Phys. Rev. Lett. {\bf 101}, 087001 (2008).

\bibitem{hchen}
H. Chen, Y. Ren, Y. Qiu, Wei Bao, R. H. Liu, G. Wu, T. Wu, Y. L.
Xie, X. F. Wang, Q. Huang and X. H. Chen, Europhys. Lett. {\bf 85},
17006(2009).

\bibitem{PCDai}
Clarina de la Cruz, Q. Huang, J. W. Lynn, Jiying Li, W. Ratcliff II,
J. L. Zarestky, H. A. Mook, G. F. Chen, J. L. Luo, N. L. Wang and
Pengcheng Dai, Nature {\bf 453} 899 (2008).

\bibitem{MKWu}
F. C. Hsu, J. Y. Luo, K. W. The, T. K. Chen, T. W. Huang, P. M. Wu,
Y. C. Lee, Y. L. Huang, Y. Y. Chu, D. C. Yan and M. K. Wu, Proc.
Nat. Acad. Sci. {\bf 105}, 14262 (2008).

\bibitem{cava}
S. Medvedev, T. M. McQueen, I. Trojan, T. Palasyuk, M. I. Eremets,
R. J. Cava, S. Naghavi, F. Casper, V. Ksenofontov, G. Wortmann and
C. Felser, Nature Mater. {\bf 8} 630(2009)

\bibitem{xlchen}
J. Guo, S. Jin, G. Wang, S. Wang, K. Zhu, T. Zhou, M. He and X.
Chen, Phys. Rev. B {\bf 82}, 180520 (2010).

\bibitem{Mizuguchi}
Yoshikazu Mizuguchi, Hiroyuki Takeya, Yasuna Kawasaki, Toshinori
Ozaki, Shunsuke Tsuda, Takahide Yamaguchi and Yoshihiko Takano,
arXiv:1012.4950 (unpublished).

\bibitem{Wang}
A. F. Wang, J. J. Ying, Y. J. Yan, R. H. Liu, X. G. Luo, Z. Y. Li,
X. F. Wang, M. Zhang, G. J. Ye, P. Cheng, Z. J. Xiang, X. H.
 Chen, arXiv:1012.5525 (unpublished).

\bibitem{Ying}
J. J. Ying, X. F. Wang, X. G. Luo, A. F. Wang, M. Zhang, Y. J. Yan,
Z. J. Xiang, R. H. Liu, P. Cheng, G. J. Ye, X. H. Chen ,
arXiv:1012.5552 (unpublished).

\bibitem{Krzton}
A. Krzton-Maziopa, Z. Shermadini, E. Pomjakushina, V. Pomjakushin,
M. Bendele, A. Amato, R. Khasanov, H. Luetkens and K. Conder,
arXiv:1012.3637 (unpublished).

\bibitem{Fang}
Minghu Fang, Hangdong Wang, Chiheng Dong, Zujuan Li, Chunmu Feng,
Jian Chen, H.Q. Yuan, arXiv:1012.5236 (unpublished).

\bibitem{Bao}
Wei Bao, Y. Qiu, Q. Huang, M. A. Green, P. Zajdel, M. R.
Fitzsimmons, M. Zhernenkov, S. Chang, Minghu Fang, B. Qian, E. K.
Vehstedt, Jinhu Yang, H. M. Pham, L. Spinu, and Z. Q. Mao, Phys.
Rev. Lett. {\bf 102}, 247001 (2009).

\bibitem{Bendele}
M. Bendele, P. Babkevich, S. Katrych, S. N. Gvasaliya, E.
Pomjakushina, K. Conder, B. Roessli, A. T. Boothroyd, R. Khasanov,
and H. Keller,  Phys. Rev. B {\bf 82}, 212504 (2010).

\bibitem{Rodriguez}
E. E. Rodriguez, C. Stock, P-Y Hsieh, N. Butch, J. Paglione, M. A.
Green, arXiv:1012.0590 (unpublished).


\bibitem{XFWang}
X. F. Wang, T. Wu, G. Wu, R. H. Liu, H. Chen, Y. L. Xie, and X. H. Chen, New Journal of Physics {\bf 11}, 045003 (2009).

\bibitem{HaihWen}
H. Q. Luo, Z. S. Wang, H. Yang, P. Cheng, X. Y. Zhu, and H. H. Wen, Supercond. Sci. Technol. {\bf 21}, 125014  (2008).

\bibitem{HaihWen1}
P. Cheng, H. Yang, Y. Jia, L. Fang, X. Y. Zhu, G. Mu, and H. H. Wen, Phys. Rev. B {\bf 78}, 134508 (2008).

\bibitem{Braithwaite}
D. Braithwaite, B. Salce1, G. Lapertot, F. Bourdarot, C. Marin, D. Aoki, and M. Hanfland, J. Phys.: Condens. Matter {\bf 21}, 232202 (2009).

\bibitem{Wang1}
H. D. Wang, C. H. Dong, Z. J. Li, S. S. Zhu, Q. H. Mao,
C. M. Feng, H. Q. Yuan, and M. H. Fang, arXiv:1101.0462
(unpublished).

\bibitem{Zhang}
Y. Zhang, L. X. Yang, M. Xu, Z. R. Ye, F. Chen, C. He, J. Jiang, B
P. Xie, J. J. Ying, X. F. Wang, X. H. Chen, J. P. Hu, and D. L.
Feng, arXiv:1012.5980 (unpublished).

\bibitem{DingH}
T. Qian, X.-P. Wang, W.-C. Jin, P. Zhang, P. Richard, G. Xu, X.Dai, Z.Fang, J.-G Guo, X.-L. Chen, and H. Ding, arXiv:1012.6017 (unpublished).


\end{thebibliography}
\end{document}